\documentclass[a4paper]{elsarticle}
\usepackage[utf8]{inputenc}
\usepackage[english]{babel}
\usepackage{amsmath}
\usepackage{graphicx}
\usepackage{xcolor}
\newcommand{\beq}{\begin{equation}}
\newcommand{\eeq}{\end{equation}}
\begin{document}
\title{The Heston stochastic volatility model with piecewise constant parameters - efficient calibration and pricing of window barrier options}
\author[lpa]{Daniel Guterding\corref{cor1}}
\ead{daniel.guterding@gmail.com}
\cortext[cor1]{Corresponding author}

\author[lpa]{Wolfram Boenkost}
\ead{wolfram.boenkost@l-p-a.com}

\address[lpa]{Lucht Probst Associates, Gro{\ss}e Gallusstra{\ss}e 9, 60311 Frankfurt am Main, Germany}

\date{\today}


\begin{abstract}
The Heston stochastic volatility model is a standard model for valuing financial derivatives, since it can be calibrated using semi-analytical formulas and captures the most basic structure of the market for financial derivatives with simple structure in time-direction. However, extending the model to the case of time-dependent parameters, which would allow for a parametrization of the market at multiple timepoints, proves more challenging. We present a simple and numerically efficient approach to the calibration of the Heston stochastic volatility model with piecewise constant parameters. We show that semi-analytical formulas can also be derived in this more complex case and combine them with recent advances in computational techniques for the Heston model. Our numerical scheme is based on the calculation of the characteristic function using Gauss-Kronrod quadrature with an additional control variate that stabilizes the numerical integrals. We use our method to calibrate the Heston model with piecewise constant parameters to the foreign exchange (FX) options market. Finally, we demonstrate improvements of the Heston model with piecewise constant parameters upon the standard Heston model in selected cases.
\end{abstract}

\maketitle

\section{Introduction}
Over the past decades, the valuation of financial derivatives using stochastic processes has become an industry standard. The prevailing methods for valuing derivatives contracts are direct Monte Carlo methods for stochastic processes and finite difference methods for partial differential equations, which can be derived from stochastic processes via the Feynman-Kac theorem. While the parameters of any financial model can in principle be calibrated to reflect market conditions using these methods, for efficiency reasons almost all models that are practically used rely on (semi-)analytical formulas for the value of simple derivatives to facilitate the calibration. Naturally, the number of models that are (semi-)analytically solvable is very limited and (semi-)analytical calibration formulas that extend the scope of available models are highly desirable.

The most important model of quantitative finance is the Black-Scholes model, which assumes a log-normal behavior of the market. Over the past decades, several stochastic volatility models have been developed to overcome the shortcomings of the Black-Scholes model~\citep{BlackScholes1973}, such as a missing smile or skew of the volatility, i.e. its underestimation of probabilities for extreme events and the missing correlation between the direction of market moves and the volatility. Among these models, the Heston stochastic volatility model~\citep{Heston1993} plays an important role, since it reproduces market smiles and skews and can be calibrated rapidly using semi-analytical formulas. 

However, to reproduce the term structure of volatility, the Heston model has to be extended to the case of time-dependent parameters. The most simple form of such time-dependence is given by piecewise constant sets of parameters assigned to subsequent time-intervals~\citep{Mikhailov2004, Elices2007, Putschoegl2010}. A piecewise constant time-dependence naturally reflects that the market quotes prices for instruments only at discrete maturities. The numerics of the semi-analytical calibration formulas for the Heston model in general has, however, been reported to be plagued by issues such as poor convergence of integrands~\citep{Mikhailov2004, Albrecher2007}. We intend to solve these issues and provide a simple and numerically stable scheme for the calibration of the Heston model and its extension to piecewise constant parameters, while preserving the semi-analytical nature of the calibration process. Of course, most of the components of our algorithm have been known previously. The aim of our paper, however, is to combine these ideas into a straightforward and easy-to-implement scheme. 

Our manuscript is organized as follows: We start the method part by introducing the method of characteristic functions~\citep{CarrMadan1999, KahlJaeckel2005} and apply it to the example case of the Black-Scholes model. We then apply the same methodology to the case of the Heston model, which we extend to the case of piecewise constant parameters. Then we modify the relevant formulas with a Black-Scholes control variate~\citep{AndersenPiterbarg2010} that suppresses oscillations in the integrands of various integrals and, hence, leads to a computationally efficient implementation. Furthermore, we explain the general calibration strategy in the case of piecewise constant parameters.

The result part of the paper illustrates the benefits of the control variate method for numerical integration. We then show calibration results for the foreign exchange (FX) options market, in particular the calibrated volatility smile within the Heston model with piecewise constant parameters. The calibrated parameters are then used to price window-barrier options that are sensitive to the term-structure of the implied volatility surface. We conclude the paper with a summary of our results.

\section{Method}
\subsection{Characteristic function for the Black-Scholes model}
We start with the derivation of the characteristic function of the Black-Scholes (BS) model~\citep{BlackScholes1973}, which illustrates the general strategy for working with characteristic functions. Furthermore, we will later use the BS characteristic function as a control variate, when we solve the Heston model. 

The stochastic process of the Black-Scholes model for the log-spot $x_t = \text{ln}[S(t)]$ is defined by the stochastic process
\beq
dx_t = \left(r_d - r_f - \frac{\sigma^2}{2}\right) dt + \sigma dW_t
,
\eeq
where $W_t$ is a Wiener process, $\sigma$ is the volatility and $r_d$ and $r_f$ are the interest rates for the domestic and foreign currency, respectively. Applying the Feynman-Kac theorem, see f.i. Ref.~\cite{Clark2010}, the partial differential equation (PDE) for options pricing within the Black-Scholes framework is given by
\beq
0 = \partial_t C + \frac{\sigma^2}{2} \partial_{xx} C + \left(r_d - r_f - \frac{\sigma^2}{2}\right) \partial_x C - r_d C ,
\label{eq:bspricing}
\eeq
where $C$ is the value of the product that we price and $t$ is the time from emission.

We define an ansatz to solve this equation for plain-vanilla call options, which is given by
\beq
C(x, \tau, K) = e^x P_1 (x, \tau, K) - e^{-r \tau} K P_2 (x, \tau, K) ,
\label{eq:bsansatz}
\eeq
where we have introduced the abbreviations $\tau = T-t$, which is the time to maturity, and $r = r_d - r_f$. The strike of the call option is denoted by $K$. The terms $P_1$ and $P_2$ can be interpreted as risk-neutral probabilities~\citep{CoxRoss1976, Heston1993}. Within the BS model these probabilities can be expressed in closed form as
\begin{subequations}
\begin{align}
P_j (x, \tau, K) &= \Phi (d_j), \quad (j=1,2) \\[2pt]
d_j &= \frac{1}{\sigma \sqrt{\tau}} \left[ x - \text{ln}(K) + \left( r + (-1)^{j-1} \frac{\sigma^2}{2} \right) \tau \right]
,
\end{align}
\label{eq:bsanalyticprob}
\end{subequations}
where $\Phi$ denotes the cumulative distribution function of the standard normal distribution. However, we will not immediately use these formulas, since we want to illustrate how to work with characteristic functions.

Therefore, we continue with calculating the derivatives of $C(K)$.
\begin{subequations}
\begin{align}
\partial_t C &= -\partial_\tau C = - e^{-x} \partial_\tau P_1 - r e^{-r \tau} K P_2 + e^{-r \tau} K \partial_\tau P_2\\[2pt]
\partial_x C &= e^x P_1 + e^x \partial_x P_1 - e^{-r \tau} K \partial_x P_2\\[2pt]
\partial_{xx} C &= e^x P_1 + 2 e^x \partial_x P_1 + e^x \partial_{xx} P_1 - e^{-r \tau} K \partial_{xx} P_2
.
\end{align}
\label{eq:ansatzderivpart1}
\end{subequations}
Substituting the ansatz (Eq.~\ref{eq:bsansatz}) into the BS pricing PDE (Eq.~\ref{eq:bspricing}) and collecting terms in $P_1$ and $P_2$, we find that these must satisfy the PDEs
\beq
\partial_\tau P_j = \frac{\sigma^2}{2} \partial_{xx} P_j + \left(r+ (-1)^{j-1}\frac{\sigma^2}{2}\right) \partial_x P_j - r_f P_j \quad (j = 1,2)
.
\label{eq:bsprobpde}
\eeq

At this point we introduce the characteristic function, which will later be used to solve the Heston model~\citep{Heston1993}. The characteristic function $f_j$ is related to the risk-neutral probabilities via 
\beq
P_j (x, \tau, K) = \frac{1}{2} + \frac{1}{\pi} \int\limits_0^\infty d\phi \, \text{Re} \Bigg[ \frac{e^{-i\phi\text{ln}(K)}f_j(x, \tau, \phi)}{i\phi} \Bigg]
.
\label{eq:charfuncgeneral}
\eeq
The characteristic function $f_j$ satisfies the same PDE as the probability $P_j$~\citep{Heston1993}. Therefore, we make an ansatz for $f_j$, which is given by
\beq
f_j (x, \tau, \phi) = \text{exp} \left(D_j (\tau, \phi) + i \phi x\right)
,
\label{eq:bsansatzcf}
\eeq
where $D_j (\tau, \phi) $ is a function that has to be determined so that the ansatz for $f_j$ truly is a solution of the pricing PDE (Eq.~\ref{eq:bsprobpde}). We now obtain the derivatives of the characteristic function.
\begin{subequations}
\begin{align}
\partial_\tau f_j &= f_j \partial_\tau D_j\\[2pt]
\partial_x f_j &= i \phi f_j \\[2pt]
\partial_{xx} f_j &= - \phi^2 f_j
\end{align}
\end{subequations}
Substituting the ansatz for $f_j$ (Eq.~\ref{eq:bsansatzcf}) in place of the probabilities $P_j$ into Eq.~\ref{eq:bsprobpde}, we end up with an ordinary differential equation (ODE) for $D_j$.
\beq
\partial_\tau D_j = \left(r + (-1)^j \frac{\sigma^2}{2}\right) i \phi - \frac{\sigma^2}{2}\phi^2 - r_f
\eeq
Integrating this ODE in time with terminal condition $D(\tau = 0, \phi) = 0$~\citep{Heston1993} we obtain the solution
\beq
D_j (\tau, \phi) = \left[ \left(r + (-1)^j \frac{\sigma^2}{2}\right) i \phi - \frac{\sigma^2}{2}\phi^2 - r_f  \right] \tau
.
\label{eq:bsDfunc}
\eeq
Using this expression together with Eqs.~\ref{eq:bsansatzcf}, \ref{eq:charfuncgeneral} and \ref{eq:bsansatz} we can in principle price plain-vanilla call options. Of course, this approach is much less efficient than directly evaluating the Black-Scholes formula, which expresses the probabilities $P_j$ in closed form. However, we will need the characteristic function of the Black-Scholes model later on to stabilize the numerics of the characteristic function in the Heston model, i.e. when using it as a control variate.

\subsection{Characteristic function for the Heston model with piecewise constant parameters}

A concise derivation of the characteristic function for the standard Heston model is given in Ref.~\cite{Rouah2013}. Here, we concentrate on a version of the model, in which the parameters of the process are time-dependent. 
\begin{subequations}
\begin{align}
dx_t &= \left(r_d - r_f - \frac{v_t}{2} \right) dt + \sqrt{v_t} dW^x_t\\[2pt]
dv_t &= \kappa (t) \left[\theta (t) - v_t \right] dt + \xi(t) \sqrt{v_t} dW^v_t\\[2pt]
dW^x_t \cdot dW^v_t &= \rho(t) dt
\end{align}
\label{eq:hestonprocess}
\end{subequations}
Here, $x$ is again the log-spot $x_t = \text{ln}[S(t)]$, $v_t$ is the instantaneous variance, $W^x_t$ and $W^v_t$ are Wiener processes, $\rho (t)$ is the correlation between those processes, $\kappa (t)$ is the speed of mean-reversion, $\theta (t)$ is the long-term variance and $\xi (t)$ is the volatility of volatility.

For $\kappa (t)$, $\theta (t)$, $\rho(t)$ and $\xi (t)$ we assume that these are piecewise constant within the time-interval $[t_i, t_{i+1})$ and that parameters undergo a discrete jump at $t_{i+1}$, where the next time-interval with constant parameters begins. In addition to these piecewise constant parameters the model also needs the initial level of variance $v_0 = v(t=0)$ as an input, which we assume to be non-time-dependent, i.e. globally constant.

Again applying the theorem of Feynman and Kac, we obtain the pricing PDE for the Heston model with piecewise constant parameters
\beq
\begin{split}
0 =& \partial_t C + \frac{v}{2} \partial_{xx} C + \frac{\xi^2 (t)}{2} v \partial_{vv} C + \xi (t) \rho (t) v \partial_{xv} C \\
 &+ \left( r_d - r_f - \frac{v}{2} \right) \partial_x C + \kappa (t) \left[\theta (t) - v\right] \partial_v C - r_d C
.
\end{split}
\label{eq:hestonpricing}
\eeq

In analogy to the previously presented BS case (Eq.~\ref{eq:bsansatz}) we use an ansatz to solve the pricing PDE.
\beq
C(x, \tau, K) = e^x P_1 (x, v, \tau, K) - e^{-r \tau} K P_2 (x, v, \tau, K)
\label{eq:hestonansatz}
\eeq
In addition to the previously calculated derivatives by $\tau$ and $x$ (see Eq.~\ref{eq:ansatzderivpart1}), we calculate the derivatives with respect to $v$.
\begin{subequations}
\begin{align}
\partial_v C &= e^x \partial_v P_1 - e^{-r\tau} K \partial_v P_2\\[2pt]
\partial_{vv} C &= e^x \partial_{vv} P_1 - e^{-r\tau} K \partial_{vv} P_2\\[2pt]
\partial_{xv} C &= e^x \partial_v P_1 + e^x \partial_{xv} P_1 - e^{-r\tau} K \partial_{xv} P_2
\end{align}
\end{subequations}
Substituting the ansatz from Eq.~\ref{eq:hestonansatz} into the pricing PDE (Eq.~\ref{eq:hestonpricing}) and collecting terms in $P_1$ and $P_2$ we find that in the Heston case these risk-neutral probabilities $P_j$, and hence the to be defined characteristic function $f_j$, must satisfy the PDEs
\beq
\begin{split}
\partial_\tau P_j =& \frac{v}{2} \partial_{xx} P_j + \frac{\xi^2 (t)}{2} v \partial_{vv} P_j + \xi (t) \rho (t) v \partial_{xv} P_j \\ &+ \left(r + (-1)^{j-1} \frac{v}{2}\right) \partial_x P_j + (a (t) - b_j (t) v) \partial_v P_j - r_f P_j
\quad (j = 1,2)
,
\end{split}
\label{eq:hestonprobpde}
\eeq
where the newly introduced coefficients $a$ and $b_j$ are given by $a(t) = \kappa (t) \theta (t) $, $b_1 (t) = \kappa (t) - \xi (t) \rho (t)$ and $b_2 (t) = \kappa (t)$.

Now we use the original ansatz for the characteristic function proposed in Ref.~\cite{Heston1993} to solve the PDEs for the risk-neutral probabilities. This ansatz is given by
\beq
f_j (x, v, \tau, \phi) = \text{exp}\left(C_j (\tau, \phi) + D_j (\tau, \phi) v + i \phi x \right)
.
\label{eq:hestonansatzcf}
\eeq
Functions $C_j (\tau, \phi)$ and $D_j (\tau, \phi)$ have to be determined so that the ansatz actually solves Eq.~\ref{eq:hestonprobpde}. To this end, we calculate the derivatives of the characteristic function.
\begin{subequations}
\begin{align}
\partial_\tau f_j &=f_j (\partial_\tau C_j + v \partial_\tau D_j)\\[2pt]
\partial_x f_j &= i \phi f_j\\[2pt]
\partial_{xx} f_j &= -\phi^2 f_j\\[2pt]
\partial_v f_j &= D_j f_j\\[2pt]
\partial_{vv} f_j &= D_j^2 f_j\\[2pt]
\partial_{xv} f_j &= i \phi D_j f_j
\end{align}
\end{subequations}
After inserting the ansatz for the characteristic function $f_j$ (Eq.~\ref{eq:hestonansatzcf}) in place of $P_j$ into Eq.~\ref{eq:hestonprobpde}, we end up with a number of terms that are linear in $v$ and others terms that do not depend on $v$. Therefore, following the original paper by Heston~\cite{Heston1993}, we write down separate ODEs for these groups of terms. Consequently, the linear factor $v$ drops out and the resulting systems of ODEs do not depend on the level of variance. Later on, when calculating the characteristic function numerically, we will set $v$ in Eq.~\ref{eq:hestonansatzcf} to $v_0 = v(t = 0)$. The set of ODEs we have to solve is now given by
\begin{subequations}
\begin{align}
\partial_\tau D_j &= \underbrace{\frac{\xi^2 (\tau)}{2}}_{= N(\tau)} D_j^2 + \underbrace{\left[\xi (\tau) \rho (\tau) i \phi - b_j (\tau)\right]}_{= - M_j (\tau, \phi)} D_j  + \underbrace{\frac{(-1)^{j-1}}{2} i\phi - \frac{\phi^2}{2}}_{= L_j (\phi)} \\[2pt]
\partial_\tau C_j &= ri\phi - r_f + a (\tau) D_j
.
\end{align}
\end{subequations}
Now we consider a time interval $[\tau_0, \tau)$ in which the parameters $\kappa (t)$, $\theta (t)$, $\rho(t)$ and $\xi (t)$ are constant. We obtain the solution of the ODE for $D_j$ and subsequently insert the solution into the ODE for $C_j$. Defining the abbreviation $A_j (\tau, \phi) = \sqrt{4 L_j (\phi) N (\tau) - M_j^2 (\tau, \phi) }$ the solution of this system of ODEs with general initial condition $D_j (\tau = \tau_0, \phi) = D_{j0}$ and $C_j(\tau = \tau_0, \phi) = C_{j0}$ can be written as
\begin{subequations}
\begin{align}
\label{eq:hestonDfunc}
D_j (\tau, \phi) &= \frac{1}{2 N} \left[ M_j + A_j \text{tan} \left( \frac{1}{2} (\tau - \tau_0) A_j + \text{arctan} \left( \frac{2 N D_{j0} - M_j}{A_j} \right) \right) \right]\\[2pt]
\begin{split}
C_j (\tau, \phi) &= C_{j0} + \frac{1}{2N} \Bigg\{ (\tau - \tau_0) \big(a M_j + 2N(r i \phi - r_f)\big)\\[2pt] 
& - a \Bigg[ \text{ln} \Bigg( 1 + \frac{(2N D_{j0} - M_j)^2}{A_j^2} \Bigg)\\[2pt]
& \quad + 2 \text{ln} \Big( \text{cos} \Big[ \frac{1}{2} A_j (\tau - \tau_0) + \text{arctan} \Big( \frac{2 N D_{j0} - M_j}{A_j} \Big) \Big] \Big) \Bigg] \Bigg\} 
.
\label{eq:hestonCfunc}
\end{split}
\end{align}
\label{eq:hestonCDfunc}
\end{subequations}
Here, we have dropped the dependencies of $A_j$, $L_j$, $M_j$ and $N$ on $\tau$ and $\phi$ for notational brevity. While the dependency on $\phi$ still exists, the dependency on $\tau$ simply means that appropriate values, constant for the time interval under consideration, have to be inserted in place of the parameters $\kappa$, $\theta$, $\rho$ and $\xi$.

Let us now assume that we are given two intervals in time to maturity $\tau$, namely $[0, \tau_1)$ and $[\tau_1, T]$, and a related set of constant parameters for each time interval, $(\kappa_0, \theta_0, \rho_0, \xi_0)$ and $(\kappa_1, \theta_1, \rho_1, \xi_1)$. To calculate the characteristic function $f_j (\tau, \phi)$ at $\tau = T$, i.e. at the time of emission $t = 0$, we start by calculating $D_j(\tau_1, \phi)$ and $C_j(\tau_1, \phi)$  according to Eq.~\ref{eq:hestonCDfunc} with initial conditions $\tau_0 = 0$ and $D_{j0} = C_{j0} = 0$ and using parameters $(\kappa_0, \theta_0, \rho_0, \xi_0)$ in place of $\big(\kappa (t),\theta (t),\rho(t),\xi (t)\big)$. Subsequently, we obtain $D_j(T, \phi)$ and $C_j(T, \phi)$ by using  Eq.~\ref{eq:hestonCDfunc} with initial conditions $\tau_0 = \tau_1$, $D_{j0} = D_j(\tau_1, \phi)$ and  $C_{j0} = C_j(\tau_1, \phi)$ and using parameters $(\kappa_1, \theta_1, \rho_1, \xi_1)$ in place of $\big(\kappa (t),\theta (t),\rho(t),\xi (t)\big)$. After completing this iteration procedure, the characteristic function can be obtained by inserting the final $D_j(T, \phi)$ and $C_j(T, \phi)$ into Eq.~\ref{eq:hestonansatzcf}. Plain-vanilla call prices can then be calculated from the characteristic function using Eqs.~\ref{eq:charfuncgeneral} and \ref{eq:hestonansatz}.

\subsection{Black-Scholes control variate method for calculating plain-vanilla call prices within the Heston model with piecewise constant parameters}
An early treatment of the Heston model with piecewise constant parameters is presented in Ref.~\cite{Mikhailov2004}. The authors reached results similar to the ones presented in the previous section, but reported numerical problems with the necessary integrals and slow convergence behavior. Therefore, we extend the formalism once more, using the idea of a control variate~\citep{AndersenPiterbarg2010} to stabilize the numerics and, in particular, to make the integral that appears in the formula for the risk-neutral probabilities (Eq.~\ref{eq:charfuncgeneral}) converge faster. 

We take the ansatz for the plain-vanilla call price within the Heston model with piecewise constant parameters (Eq.~\ref{eq:hestonansatz}) and add zero in the form of the Black-Scholes price of the same plain-vanilla call option calculated using the direct Black-Scholes formula for the risk-neutral probabilities $P_j$ subtracted by the Black-Scholes price expressed through the characteristic function (using Eqs.~\ref{eq:charfuncgeneral}, \ref{eq:bsansatzcf} and \ref{eq:bsDfunc}). For the Black-Scholes pricing we assume $\sigma = \sqrt{v_0}$. This results in a modified formula for the plain-vanilla call price within the Heston model with piecewise constant parameters.
\beq
\begin{split}
C(K) &= C^{BS} (K) +  e^x \left[ P_1^H  - P_1^{BS}\right] (x, v, \tau, K) - e^{-r \tau} K \left[ P_2^H - P_2^{BS} \right] (x, v, \tau, K)  \\[2pt]
&= C^{BS} (K) + e^x \tilde P_1 (x, v, \tau, K)  - e^{-r \tau} K \tilde P_2 (x, v, \tau, K) 
\end{split}
\label{eq:hestonptildepricing}
\eeq
The quantity $\tilde P_j$ is the difference of the risk-neutral probabilities between the Heston model with piecewise constant parameters and the Black-Scholes model. Hence, $\tilde P_j$ itself is not a probability and can assume negative values. Using Eq.~\ref{eq:charfuncgeneral} for the risk-neutral probability expressed through the characteristic function $f_j$ we get an expression for $\tilde P_j$.
\beq
\tilde P_j (x, v, \tau, K) = \frac{1}{\pi} \int\limits_0^\infty d \phi \, \text{Re} \left[ \frac{e^{-i \phi \text{ln} (K)}}{i \phi} \left( f_j^H (x, v, \tau, K) - f_j^{BS} (x, \tau, K) \right) \right]
\label{eq:hestonptilde}
\eeq
Here, $f_j^{BS}$ is the characteristic function of the Black-Scholes model defined by Eq.~\ref{eq:bsansatzcf}, while $f_j^H$ is the characteristic function of the Heston model with piecewise constant parameters defined by Eq.~\ref{eq:hestonansatzcf}. In all BS terms we use $\sigma = \sqrt{v_0}$.

The crucial point about the control variate approach is that $C^{BS}(K)$ is calculated from the closed expression for the risk-neutral probabilities (Eqs.~\ref{eq:bsanalyticprob} and \ref{eq:bsansatz}). Therefore, $\tilde P_j$ just contains corrections of the piecewise constant Heston model with respect to the Black-Scholes model.

In practice, we calculate $\tilde P_j$ using the computationally efficient Gauss-Kronrod quadrature~\citep{Kahaner1989}. Note that this way we automatically avoid the evaluation of the integrand in Eq.~\ref{eq:hestonptilde} at the numerically problematic lower boundary ($\phi = 0$), since the quadrature nodes never coincide with the boundary points.

\subsection{Calibration of the Heston model with piecewise constant parameters to the foreign exchange options market}
For the calibration of the model defined by Eq.~\ref{eq:hestonprocess} we use a range of plain-vanilla call options with different strikes and maturities. Since FX options are conventionally quoted with respect to the option delta, we use five strikes equivalent to an option delta of $\Delta = (\pm 0.15, \pm 0.25, 0.5)$. The calibration of the Heston parameters is facilitated by numerically optimizing the option prices generated from Eq.~\ref{eq:hestonptildepricing} with respect to the market price using the Levenberg-Marquardt algorithm~\citep{Press2007}.

As a starting point for finding piecewise constant parameters, we first numerically optimize a global parameter set for the entire time interval $[0,T_N]$ up to the longest maturity of interest $T_N$. In this first optimization step the speed of mean-reversion is kept constant at $\kappa = 1.5$. After finding the globally optimal parameters, the initial variance $v_0$ is fixed to the value found in the global optimization step. The values for $\theta$, $\kappa$, $\rho$ and $\xi$ are used as initial values for the optimization when attempting to find time-dependent parameters.

The time-dependence of parameters $\theta$, $\kappa$, $\rho$ and $\xi$ is now introduced by bootstrapping parameter sets from the shortest to the longest maturity. We start with the shortest maturity $T_0$ and numerically optimize the parameters, which we then assign to the time interval $[0, T_0)$. Then we proceed to the next longer maturity $T_1$ and determine the parameters for the time interval $[T_0, T_1]$. We continue the iteration until parameters for all desired time intervals are determined.

Note that within the optimization process bounds must be imposed on the parameters. The correlation $\rho$ is limited to the interval $[-1:1]$, while the other parameters in general have to be positive. In practice, we impose bounds also on the other parameters to prevent the optimization algorithm from exploring regimes of excessively small or large parameters.

\section{Results}
\subsection{Effect of Black-Scholes control variate on integrands for risk-neutral probabilities}
Here, we show results for the integrand of the characteristic function of plain-vanilla call options in various settings to demonstrate the advantages of using a Black-Scholes control variate. The parameter sets we use are given in terms of a standard Heston model with constant parameters and were calibrated to the market on May 23rd, 2017. The calibration results are given in Table~\ref{tab:hestonparameterssimple}.

\begin{table}[bt]
\setlength\tabcolsep{10pt}
\caption{Plain Heston model parameters calibrated to market-quoted plain-vanilla call options on May 23rd, 2017.}
\begin{tabular}{r r r r r r r}
ccy pair & T & $v_0 \times100$ & $\theta \times10$ & $\kappa$ & $\rho$ & $\xi$ \\
\hline
EUR/USD & 2M & 0.675 & 0.199 & 1.04 & -0.276 & 0.313 \\
EUR/USD & 1Y & 0.674 & 0.166 & 0.92 & -0.239 & 0.190 \\ 
EUR/GBP & 2M & 0.697 & 0.177 & 1.09 & -0.032 & 0.290 \\
EUR/GBP & 1Y & 0.648 & 0.104 & 0.79 & 0.187 & 0.174 \\
EUR/JPY & 2M & 0.944 & 0.175 & 1.40 & -0.403 & 0.257 \\ 
EUR/JPY & 1Y & 0.893 & 0.148 & 0.89 & -0.265 & 0.240 \\
\end{tabular}
\label{tab:hestonparameterssimple}
\end{table}

\begin{figure*}[t]
\includegraphics[width=\linewidth]{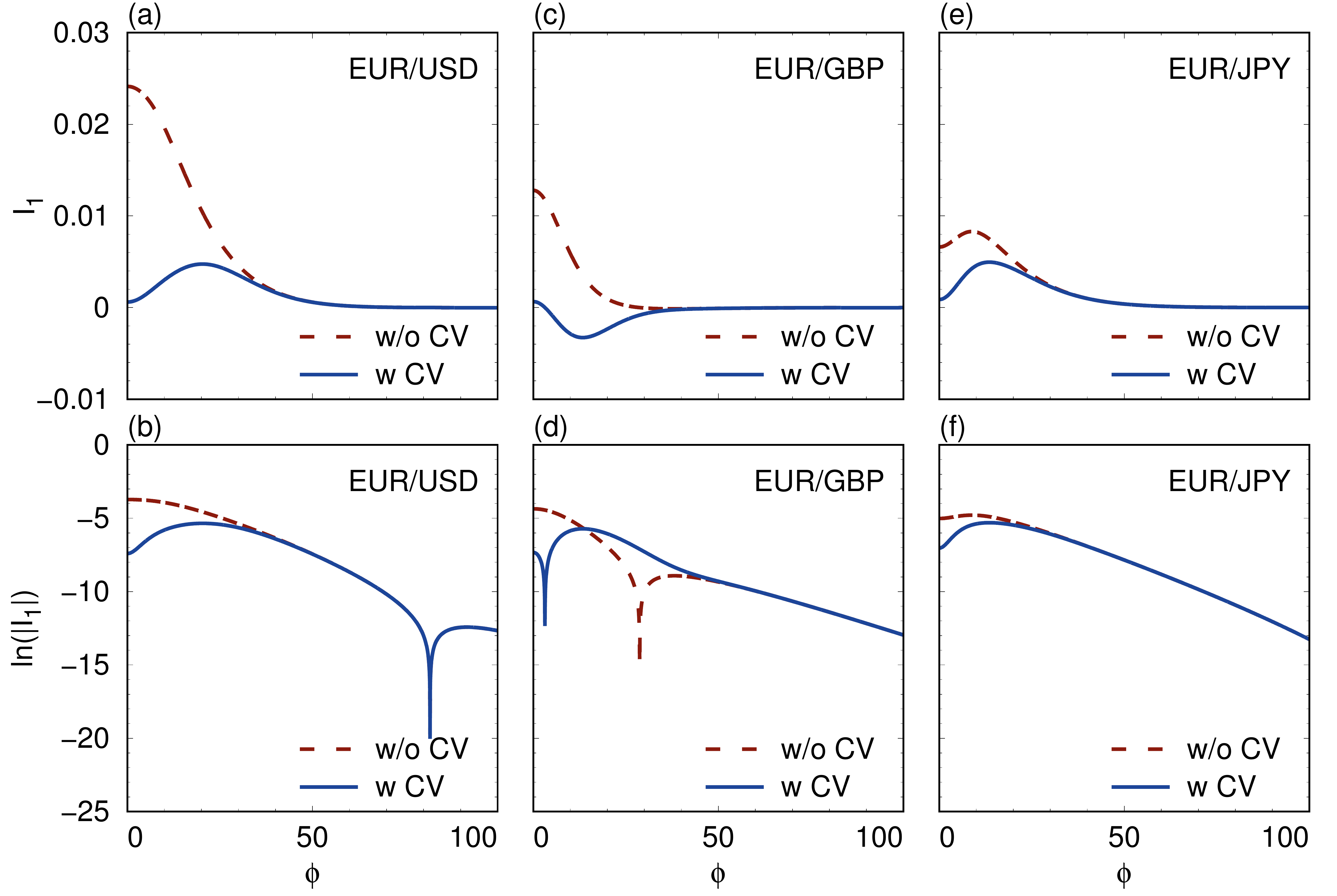}
\caption{(Color online) Integrand (top row) and logarithm of the absolute value of the integrand (bottom row) of Eq.~\ref{eq:hestonptilde} without (w/o) and with (w) control variate (CV) as a function of expansion parameter $\phi$ for an at-the-money plain-vanilla call option in the Heston model. Parameters for maturity $T$=1Y are taken from Table~\ref{tab:hestonparameterssimple}.}
\label{fig:controlvar1y}
\end{figure*}

In Fig.~\ref{fig:controlvar1y} we show the integrand of Eq.~\ref{eq:hestonptilde} and the logarithm of its absolute value with and without control variate plotted against the expansion parameter $\phi$ for an at-the-money call option with expiry after one year. The integrand is generally well-behaved and easy to integrate both with and without the control variate approach. Although the control variate approach reduces the magnitude of the integrand at the origin ($\phi = 0$) by several orders, it can lead to increases at larger $\phi$ as seen in the EUR/GBP case (see Fig.~\ref{fig:controlvar1y}(c,d)). When pricing at-the-money call options we did not find any numerical difficulties, even without the control variate approach. Compared to literature results~\citep{Mikhailov2004} this may be due to our slightly different formulas for the solution of the Heston Riccati ODEs (see Eq.~\ref{eq:hestonCDfunc}). 

\begin{figure*}[t]
\includegraphics[width=\linewidth]{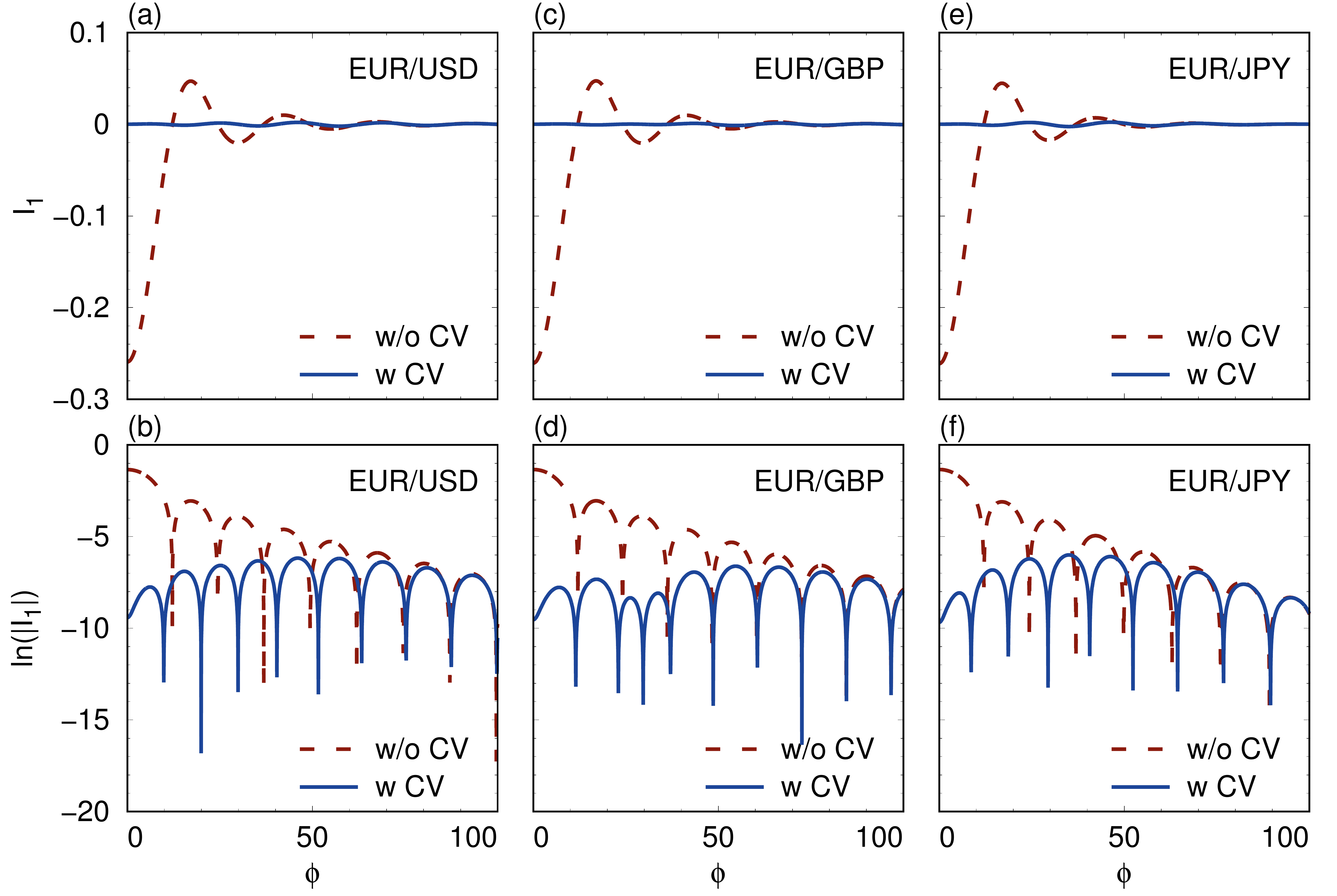}
\caption{(Color online) Integrand (top row) and logarithm of the absolute value of the integrand (bottom row) of Eq.~\ref{eq:hestonptilde} without (w/o) and with (w) control variate (CV) as a function of expansion parameter $\phi$ for an out-of-the-money plain-vanilla call option with strike $K = 1.3 S_0$ in the Heston model. Parameters for maturity $T$=2M are taken from Table~\ref{tab:hestonparameterssimple}.}
\label{fig:controlvar2m}
\end{figure*}

The power of the control variate approach is more evident when pricing options that are far out-of-the-money, where the integrand without control variate is much less well-behaved. In Fig.~\ref{fig:controlvar2m} we show results for the integrand of Eq.~\ref{eq:hestonptilde} for an out-of-the-money call option with strike $K = 1.3 S_0$. Without control variate, the integrand displays wild oscillations that are damped with increasing expansion parameter $\phi$. Using a control variate reduces the magnitude of the integrand at the origin by about eight orders and strongly suppresses the magnitude of oscillations. This results in a vast advantage when it comes to numerical integration, due to a significantly reduced number of function evaluations.

Note, however, that our far out-of-the-money example represents a somewhat extreme case that is not likely to occur in real-world applications. Nevertheless, it demonstrates that the control variate approach is a simple way to stabilize the numerics of the Heston model when pricing options that involve less well-behaved integrands.

\subsection{Calibration of the Heston model with piecewise constant parameters to the term-structure of the foreign exchange options market}
In the following we want to price EUR/USD window barrier options with times to maturity 1M, 3M, 6M and 1Y. Therefore, we need to calibrate Heston models for those maturities. We subdivide each of the intervals into three subintervals (four in the case of 1Y to maturity) and calibrate piecewise parameters for each sub-interval using the methodology described above. The resulting parameter sets are shown in Table~\ref{tab:hestonparameterspiecewise}. The set indices (1,2,3,4) correspond to the times to maturity (1M, 3M, 6M, 1Y). 

\begin{table}[bt]
\setlength\tabcolsep{10pt}
\caption{Piecewise Heston model parameters calibrated to market-quoted EUR/USD plain-vanilla call options on August 7th, 2013.}
\begin{tabular}{r r r r r r r r}
Set Index & From & To & $v_0 \times100$ & $\theta \times10$ & $\kappa$ & $\rho$ & $\xi$ \\
\hline
1 & 0 & 1W & 0.7 & 0.00 & 2.986 & -0.064 & 0.771 \\ 
 & 1W & 3W & 0.7 & 0.29 & 6.000 & -1.000 & 0.484  \\
 & 3W & 1M & 0.7 & 0.00 & 6.000 & -0.209 & 0.681 \\
\hline
2 & 0 & 1M & 0.6 & 0.13 & 1.539 & -0.322 & 0.281\\
 & 1M & 2M & 0.6 & 0.12 & 4.948 & -0.764 & 0.173\\
 & 2M & 3M & 0.6 & 0.02 & 1.735 &  -0.288 & 0.367\\
\hline
3 & 0 & 2M & 0.6 & 0.09 & 1.859 & -0.374 & 0.232 \\
 & 2M & 4M & 0.6 & 0.12 & 3.025 & -0.549 & 0.310 \\
 & 4M & 6M & 0.6 & 0.05 & 2.855 & -0.477 & 0.203 \\
\hline
4 & 0 & 1M & 0.6 & 0.10 & 1.443 & -0.321 & 0.277\\
 & 1M & 2M & 0.6 & 0.12 & 4.985 & -0.693 & 0.198\\
 & 2M & 6M & 0.6 & 0.07 & 2.430 & -0.437 & 0.268\\
 & 6M & 1Y & 0.6 & 0.16 & 1.613 & -0.503 & 0.328\\
\end{tabular}

\label{tab:hestonparameterspiecewise}
\end{table}

The stability of the calibration procedure is illustrated by the 0 to 1M sub-intervals in sets 2 and 4, which start the piecewise optimization from different globally optimal parameter sets, but should arrive at the same calibrated parameters if the algorithm is stable, since they calibrate to the same data. We observe that the deviations between different optimization runs are tiny and that the optimization is in general stable (see Table~\ref{tab:hestonparameterspiecewise}). For shorter maturities (sets 1 and 2) the quality of calibration is not very good and calibrated parameters are sometimes close to the imposed bounds ($\rho \in [-1:1]$, $\kappa \in (0, 6]$). The upper bound for $\kappa$ is somewhat artificial, but prevents the optimization algorithm from exploring excessively large speeds of mean-reversion, which can lead to slow calibration. This indicates that using very short time-intervals smaller than one month in the piecewise constant Heston model is not advisable.

\begin{figure*}[t]
\includegraphics[width=\linewidth]{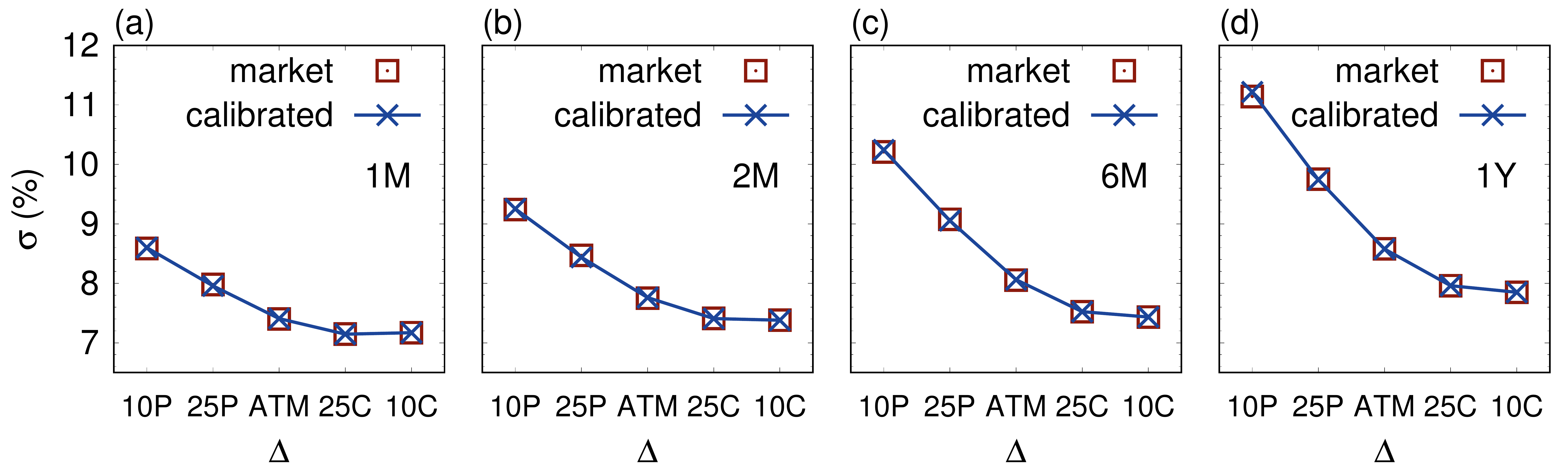}
\caption{(Color online) Calibrated implied volatilities within the piecewise Heston model for EUR/USD plain-vanilla call options on August 7th, 2013. Note the strong term-structure of implied volatility, which increases from short maturities to longer ones. The values of volatility are given in percent and quoted versus the Delta of plain-vanilla put (10P, 25P) and plain-vanilla call options (25C, 10C). ATM stands for the at-the-money Delta. Lines serve as guides to the eye. The implied volatilities shown here correspond to parameter set No.~4 in Table~\ref{tab:hestonparameterspiecewise}.}
\label{fig:calibration1y}
\end{figure*}

For parameter set 4, which corresponds to an overall product time to maturity of 1Y, we show the calibrated and market-quoted implied volatilities for each time-slice in Fig.~\ref{fig:calibration1y}. Note the excellent quality of calibration despite the strong term structure of volatility contained in the market, which increases from short to longer maturities. In such cases, the Heston model with piecewise constant parameters is able to reproduce the market and improvements over a plain Heston model calibrated against maturity can be expected for products that are sensitive to the term-structure of volatility. Therefore, we continue with the pricing of window barrier options.

\subsection{Pricing of window-barrier options within the Heston model with piecewise constant parameters}
Recently, Tian et al.~\cite{Tian2013} have presented a methodology to price window barrier options using a stochastic local volatility model, which shows good agreement with market prices. The reference prices contained in Ref.~\cite{Tian2013} correspond to actual traded options. Here, we cite the parameters of these options and the reference prices (Table~\ref{tab:windowbarrierspecs}).

\begin{table}[bt]
\caption{Window barrier options with reference prices taken from Ref.~\cite{Tian2013}. In the trigger column, L corresponds to a lower barrier, while U corresponds to an upper barrier.}
\setlength\tabcolsep{10pt}
\renewcommand{\arraystretch}{0.7}
\scalebox{0.7}{
\begin{tabular}{r r r r r r r r}
 Maturity & Index  & Trigger & Barrier start & Barrier end & Reference & H & HPW \\ \hline
 1m, 2013/09/05 & 1 & L=1.26 &  2013/08/07 &  2013/08/21 & 0.0002 & 0.0005 & 0.0006 \\ 
& 2 & L=1.26 & 2013/08/21 & 2013/09/05 & 0.0017  & 0.0025 & 0.0024   \\ 
& 3 & L=1.26 & 2013/08/07 & 2013/09/05 & 0.0017 & 0.0025 & 0.0024 \\ 
& 4 & U=1.36 & 2013/08/07 & 2013/08/21 & 0.0036 & 0.0043 & 0.0043 \\ 
&5 & U=1.36 & 2013/08/21 & 2013/09/05 & 0.0079 & 0.0080 &  0.0081\\ 
&6 & U=1.36 & 2013/08/07 & 2013/09/05 & 0.0080 & 0.0081 & 0.0081 \\ \hline
3m, 2013/11/08 & 7 & L=1.23 & 2013/08/07 & 2013/08/21 & 0.0000 & 0.0003 & 0.0004 \\ 
&8 & L=1.23 & 2013/08/07 & 2013/11/01 & 0.0083 & 0.0093 & 0.0093 \\ 
&9 & L=1.23 & 2013/08/21 & 2013/11/08 & 0.0094 & 0.0100 & 0.0100 \\ 
&10 & L=1.23 & 2013/11/01 & 2013/11/08 & 0.0080 & 0.0085 & 0.0084 \\ 
&11 & L=1.23 & 2013/08/07 & 2013/11/08 & 0.0094 & 0.0100 & 0.0100 \\ 
&12 & U=1.39 & 2013/08/07 & 2013/08/21 & 0.0002 & 0.0012 & 0.0011 \\ 
&13 & U=1.39 & 2013/08/07 & 2013/11/01 & 0.0138 & 0.0134 & 0.0133 \\ 
&14 & U=1.39 & 2013/08/21 & 2013/11/08 & 0.0148 & 0.0140 & 0.0138 \\ 
&15 & U=1.39 & 2013/11/01 & 2013/11/08 & 0.0129 & 0.0130 & 0.0128 \\ 
&16 & U=1.39 & 2013/08/07 & 2013/11/08 & 0.0148 & 0.0140 & 0.0139 \\ 
\hline
6m, 2014/02/06 & 17 & L=1.20  & 2013/08/07 & 2013/08/21  &0.0000 & 0.0003 & 0.0002 \\ 
& 18 & L=1.20 & 2013/08/07 & 2013/11/01 & 0.0043 & 0.0056 & 0.0055 \\ 
& 19 & L=1.20 & 2013/08/07 & 2014/01/31 & 0.0159 & 0.0170 & 0.0171 \\ 
& 20 & L=1.20 & 2013/08/21 & 2014/02/06 & 0.0165 & 0.0170 & 0.0171 \\ 
& 21 & L=1.20 & 2013/11/01 & 2014/02/06 & 0.0164 & 0.0168 & 0.0168 \\ 
&22 & L=1.20  & 2014/01/31 & 2014/02/06 & 0.0135 & 0.0136 & 0.0136 \\ 
&23 & L=1.20 & 2013/08/07 & 2014/02/06 & 0.0165 & 0.0170 & 0.0171 \\ 
&24 & U=1.42 & 2013/08/07 & 2013/08/21 & 0.0000 & 0.0005 & 0.0004 \\ 
&25 & U=1.42 & 2013/08/07 & 2013/11/01 & 0.0065 & 0.0078 & 0.0076 \\ 
&26 & U=1.42 & 2013/08/07 & 2014/01/31 & 0.0189 & 0.0185 & 0.0183 \\ 
&27 & U=1.42 & 2013/08/21 & 2014/02/06 & 0.0195 & 0.0185 & 0.0183 \\ 
&28 & U=1.42 & 2013/11/01 & 2014/02/06 & 0.0193 & 0.0183 & 0.0181 \\ 
&29 & U=1.42 & 2014/01/31 & 2014/02/06 & 0.0160 & 0.0163 & 0.0161 \\ 
&30 & U=1.42 & 2013/08/07 & 2014/02/06 & 0.0195 & 0.0185 & 0.0183 \\ \hline
1y, 2014/08/07 & 31 & L=1.15 & 2013/08/07 & 2013/08/21 & 0.0000  & 0.0002 & 0.0002 \\
& 32 &  L=1.15 & 2013/08/07 & 2013/11/01 & 0.0009  & 0.0023 & 0.0020 \\
&33 &  L=1.15 & 2013/08/07 & 2014/01/31 & 0.0085  & 0.0111 & 0.0099 \\
& 34 & L=1.15 & 2013/08/07 & 2014/07/07 & 0.0253  & 0.0260 & 0.0258 \\
&35 & L=1.15 & 2013/08/21 & 2014/07/07 & 0.0253  & 0.0260 & 0.0258 \\
&36 & L=1.15 & 2013/11/01 & 2014/07/07 & 0.0253 & 0.0259 & 0.0257 \\
&37 & L=1.15 & 2014/01/31 & 2014/07/07 & 0.0250  & 0.0253 & 0.0251 \\
&38 & L=1.15 & 2013/08/07 & 2014/08/07 & 0.0282  & 0.0282 & 0.0285 \\
&39 & U=1.45 & 2013/08/07 & 2013/08/21 & 0.0000 & 0.0002 & 0.0002 \\ 
&40 & U=1.45 & 2013/08/07 & 2013/11/01 & 0.0024 & 0.0044 & 0.0038 \\ 
&41 & U=1.45 & 2013/08/07 & 2014/01/31 & 0.0117 & 0.0142 & 0.0123 \\ 
&42 & U=1.45 & 2013/08/07 & 2014/07/07 & 0.0314 & 0.0293 & 0.0285 \\ 
&43 & U=1.45 & 2013/08/21 & 2014/07/07 & 0.0314 & 0.0293 & 0.0285 \\ 
&44 & U=1.45 & 2013/11/01 & 2014/07/07 & 0.0313 & 0.0292 & 0.0285 \\ 
&45 & U=1.45 & 2014/01/31 & 2014/07/07 & 0.0310 & 0.0288 & 0.0281 \\ 
&46 & U=1.45 & 2013/08/07 & 2014/08/07 & 0.0345 & 0.0315 & 0.0312 \\ 
\end{tabular}
}
\label{tab:windowbarrierspecs}
\end{table}

\begin{figure*}[t]
\includegraphics[width=\linewidth]{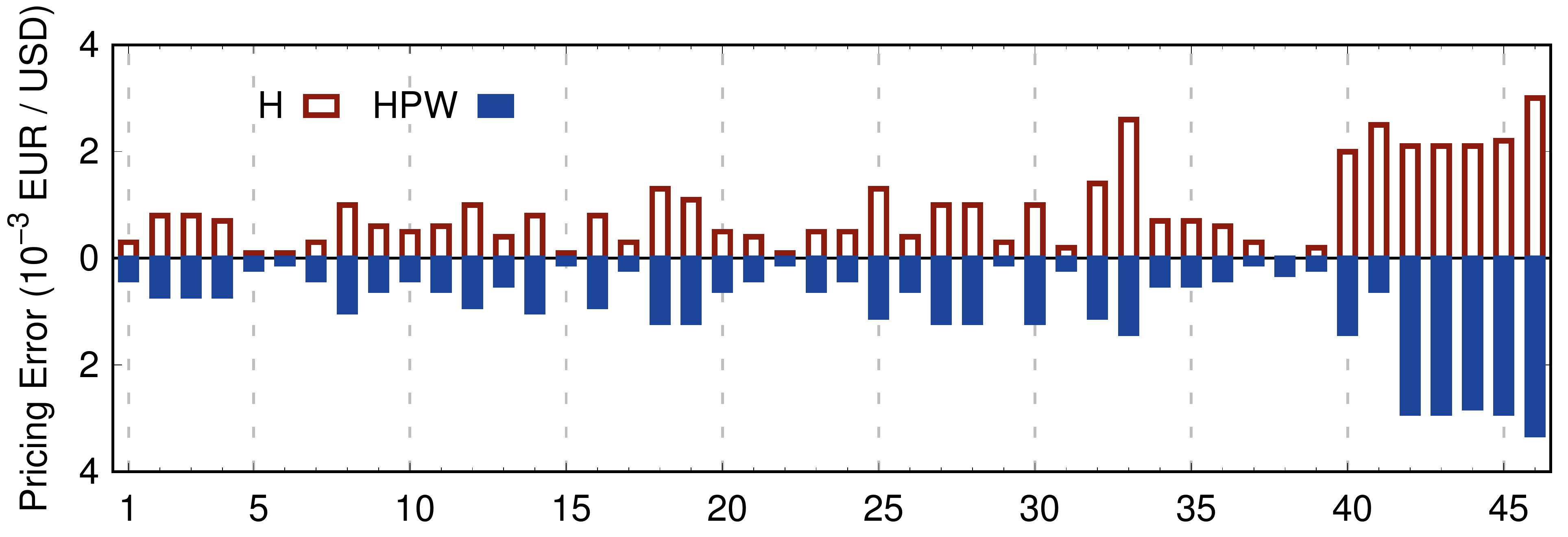}
\caption{(Color online) Pricing error within the plain Heston model (H) and the Heston model with piecewise constant parameters (HPW) plotted against the index of the priced window barrier option. The option prices are calculated in EUR per USD.}
\label{fig:barrierprices}
\end{figure*}

We do not use a stochastic local volatility model, but price these options with a plain Heston model (abbreviated H) calibrated against maturity and using the Heston model with piecewise constant parameters (abbreviated HPW) as given in Table~\ref{tab:hestonparameterspiecewise}. The option pricing is done using the finite difference method, see f.i. Ref.~\cite{Clark2010}. The prices calculated using our implementations are given in Table~\ref{tab:windowbarrierspecs}. A comparison between the Heston model with and without time-dependence of parameters is given in Fig.~\ref{fig:barrierprices}.

The prices in the Heston model with piecewise constant parameters are generally consistent with those of a plain Heston model. Significant differences in price are only visible for options with a maturity of 6M or longer, since only then the term-structure of volatility actually takes effect. Strong improvements are visible for options with indices 33, 40 and 41 (see Table~\ref{tab:windowbarrierspecs}). All of those contain short barrier windows right at the beginning of the product term, which makes them especially sensitive to the much lower volatility in this time span compared to the rest of the product lifetime. Therefore, the plain Heston model calibrated against maturity uses a too large volatility at the beginning of the product term and overestimates the probability of the knock-in barrier being reached. Accordingly, it overestimates the value of the product.

In general, however, introducing piecewise constant parameters into the Heston model only leads to improvements, where products are very sensitive to the term-structure of volatility. Other problems of the Heston model, such as poor pricing for far out-of-the money barriers, like for option numbers 42 to 46, are not cured by the piecewise constant parameters. In such cases, a hybrid stochastic local volatility model has to be considered~\citep{Clark2010, Tian2013, vanDerStoep2014, Wyns2016}.

\section{Conclusions}
We extended the calibration of the Heston stochastic volatility model via the method of characteristic functions to the case of piecewise constant parameters. In the numerical treatment of the resulting formulas we introduced the method of a control variate to suppress numerical instabilities. In combination with Gauss-Kronrod quadrature, this leads to a fast and reliable calibration strategy, even in the case of piecewise constant parameters.

We benchmarked the piecewise constant Heston model using window barrier options, which are sensitive to the term structure of volatility. For less sensitive options, the piecewise calibration does not improve upon a plain Heston model. However, we find that for maturities longer than 6M, the piecewise calibration leads to improved pricing of window barrier options with strong dependence on the term structure of volatility. In such cases, the Heston model with piecewise constant parameters in combination with our formalism for calibration offers rapid and reliable calibration, while the complexity of pricing is on the same level as for the standard Heston model.



\end{document}